# Selectively Controlled Magnetic Microrobots with Opposing Helices

Joshua Giltinan,[1] Panayiota Katsamba,[2] Wendong Wang,[1] Eric Lauga,[3] and Metin Sitti[1, 4, a)]

[1)]*Physical Intelligence Department, Max Planck Institute for Intelligent Systems, Stuttgart, 70569,
Germany*
[2)]*School of Mathematics, University of Birmingham, Edgbaston, Birmingham, B15 2TT,
United Kingdom*
[3)]*Department of Applied Mathematics and Theoretical Physics, University of Cambridge, Cambridge CB3 0WA,
United Kingdom*
[4)]*School of Medicine and School of Engineering, Koç University, 34450 Istanbul, Turkey*

Magnetic microrobots that swim through liquid media are of interest for minimally invasive medical procedures, bio-engineering, and manufacturing. Many of the envisaged applications, such as micromanipulation and targeted cargo delivery, necessitate the use and adequate control of multiple microrobots, which will increase the velocity, robustness, and efficacy of a procedure. While various methods involving heterogeneous geometries, magnetic properties, and surface chemistries have been proposed to enhance independent control, the main challenge has been that the motion between all microswimmers remains coupled through the global control signal of the magnetic field. Katsamba and Lauga proposed transchiral microrobots, a theoretical design with magnetized spirals of opposite handedness. The competition between the spirals can be tuned to give an intrinsic nonlinearity that each device can function only within a given band of frequencies. This allows individual microrobots to be selectively controlled by varying the frequency of the rotating magnetic field. Here we present the experimental realization and characterization of transchiral micromotors composed of independently driven magnetic helices. We show a swimming micromotor that yields negligible net motion until a critical frequency is reached and a micromotor that changes its translation direction as a function of the frequency of the rotating magnetic field. This work demonstrates a crucial step towards completely decoupled and addressable swimming magnetic microrobots.

Microrobots, untethered mobile machines capable of navigating and manipulating in a sub-millimeter environment, are envisioned as a technology that will revolutionize healthcare, bioengineering, and manufacturing.[1–5] For these applications, their manipulation in fluidic environments is of great interest for both applications and scientific studies. The control of multiple microrobots could increase their efficacy in various tasks, such as micromanipulation[6] and cargo delivery.[7] Thus, it is advantageous to study methodologies to control multiple magnetic microrobots.[8–12] As microrobots become smaller, nonreciprocal swimming becomes a scalable mode of propulsion at low Reynolds numbers.[13] Generating nonreciprocal motion with microrobots in low Reynolds number environments has been a topic of recent research; helical structures,[14] swimming sheets,[15–17] undulatory robots,[18] and irregularly shaped clusters[19,20] in a time-varying magnetic field have been proposed as fluidic propulsion solutions. Magnetic fields are of interest due to their long range and ability to safely penetrate tissues.[5] The magnetic field induces a magnetic torque on the swimmer, yielding a propulsive force that scales more favorably than magnetic gradient pulling.[21]

In two dimensions (2D), specialized surfaces are able restrict the motion of microrobots, such that their response to the global control signal can be individualized.[12,22,23] However, these methodologies are unable to be adapted for a workspace far from any boundaries. In three dimensions (3D), helical magnetic microrobots typically swim by rotating the magnetic field perpendicular to the desired axis of propulsion. An asymmetry in the fluid drag yields a forward force as a result of the net viscous drag on the structure. In a given frequency range, this motion leads to a stable forward propulsive force.[24,25] An example response, called the step-out profile, is illustrated in Fig. 1A.[26] If the handedness of the spiral is reversed, the swimmer will have a negative propulsion in the same rotating magnetic field, as illustrated in Fig. 1B. In each example the swimming velocity linearly increases with the frequency until a critical "step-out" frequency is reached. This is the frequency where fluid resistive torque exceeds the maximum possible magnetic torque, thus it is a function of fluid viscosity, helix drag coefficient, magnetic moment of the helix, and magnetic field magnitude. After this frequency, the velocity nonlinearly decreases to zero. Assuming the viscosity of the fluid is constant, decreasing the drag on the swimmer by altering its surface chemistry would increase its step-out frequency. Wang et al. exploited this by changing the surface chemistry on the outside of the spiral using various thiol and thiolether-based compounds. By driving more hydrophobic spirals at higher frequencies, separate groups of spirals with otherwise homogeneous geometry could be controlled.[27] As the constant of proportionality between forward velocity and magnetic field frequency is related to the geometric and magnetic properties of the spiral, control schemes can exploit the different motion primitives of swimmers to drive heterogeneous swimmers to unique locations.[28–32] Helical motors are also the smallest, mass-produced magnetic microrobots. Glancing-angle deposition (GLAAD) is able to achieve helices on the nanometer-scale,[33] and independent control of these has been demonstrated by exploiting their interactions with a nearby surface.[34] However, swimming helical microrobots currently cannot exhibit two features. The first is a zero net translation response to a rotating magnetic field. Control of heterogeneous swimming microrobots would benefit from banded frequency responses, where ideally there would be no net motion if the magnetic field was actuating outside of the frequency band. Second, reversible motion without changing the magnetic field rotation direction has currently only been demonstrated by randomly self-assembled non-helical magnetic propellers.[35]

---

[a)]Electronic mail: sitti@is.mpg.de

As the velocity profile of a single magnetised helix lacks the cutoff at low frequencies, Katsamba and Lauga proposed to couple two helices of opposite handedness in what is called the transchiral helical micromotor to achieve a banded velocity profile.[36] This interval of actuating frequencies, an effective frequency band, would allow selective control over multiple micromotors. The helices are coupled such that they can freely rotate about the axial rod but are constrained to move at the same velocity, pushing or pulling each other in the opposite direction. The geometric and magnetic properties of the two helices can be selected to tune this force balance to give rise to the required banded velocity profile. An example is illustrated in Fig. 1C. Here, before any of the two helices of the transchiral motor step-out, the force balance is such that the motor is stationary. After the first step-out frequency, the helix that has not stepped out dominates, with a monotonically increasing velocity profile, until it also steps out, after which the velocity decreases to zero as the frequency is increased further. The force balance in a transchiral motor can also be tuned to give a velocity profile with positive and negative values in different frequency ranges, as illustrated in Fig. 1D. This allows for reversal of the direction of motion by changing the actuating frequency. With different micromotors having distinct non-overlapping effective frequency bands, one can choose which to operate by tuning the magnetic field frequency appropriately. If one wishes to combine both features of selective control via banded velocity profile and reversal of motion, then at least three helices would be required.[36]

Here we present the characterization of transchiral micromotors composed of independently driven magnetic helices. In order to couple the translation of the spirals but not their rotation, we used an axial rod that passes through the central axis of the helices and has disk-shaped tapers that prevent the helices from exiting the structure. This allows freedom of rotation between the helices. The helices push against the tapers and transmit their propulsive force to the axial rod, thereby resulting in the push/pull relation explained by Katsamba and Lauga.[36] Without rotational coupling, the spirals respond to a rotating magnetic field as if they were not in the presence of other structures. As our Reynolds numbers are on the order of $10^{-2}$, we assume the spirals operate in the Stokes flow regime, and the force exerted on the passive frame by each spiral will be proportional to the swimming velocity of the spiral. We fabricate and characterize two configurations of transchiral motors. The first configuration consists of two spirals which possess homogeneous geometry, opposite handedness, and differing magnetic strengths. At low frequencies, the net propulsion should be zero, as illustrated in Fig. 1C. The second configuration consists of two helices with heterogeneous geometry, and the propulsion direction is dependent on the rotation frequency, as illustrated in Fig. 1D. The assumption of this working principle is that there is a continuous mechanical contact between each spiral and the rod, such that each spiral is exerting a force on the rod that is completely in the direction of propulsion (the long axis of the rod). Thus, by pushing against the tapered ends of the rod, the two spirals are effectively pushing or pulling each other. If the spiral is not pushing perfectly in the direction of the long axis, a portion of the transmitted force could be perpendicular to the long axis of the micromotor, which may lead to a variation in the resulting speed.

FIG. 1. **Transchiral motor concept.** Illustrations of each motor are given above the respective plots with right and left handedness marked "R" and "L". A) The characteristic velocity-frequency response of a magnetic chiral swimmer. The velocity is linearly proportional to the rotation frequency of the magnetic field, $\vec{B}$, until the step-out frequency $\Omega_s$. B) When the handedness is opposite, in the same rotating magnetic field, the spiral will propel in the opposite direction. Coupling the translation forces of two helices results in non-linear behaviour before all spirals reach their step-out frequencies. C) The velocity-frequency response of two helices with opposing handedness but otherwise identical geometry and magnetic strength. Until the magnetically weaker helix steps-out, there is no net translation. After the first spiral steps out, $\Omega_{s1}$, the velocity nonlinearly increases until the second spiral steps out at frequency $\Omega_{s2}$. D) The velocity-frequency response of two helices with heterogeneous geometry, one has a spiral diameter of 100 $\mu m$ and the other has a diameter of 200 $\mu m$. A spiral with a larger radius has a higher velocity, but a lower step-out frequency, $\Omega_{s3}$, than a spiral with a smaller radius, $\Omega_{s4}$.

TABLE I. The properties of the individual spiral species used in physical experiments.

| spiral species and handedness | wire diameter ($\mu m$) | length ($\mu m$) | turns | spiral diameter ($\mu m$) |
|---|---|---|---|---|
| $A_L$ | 20 | 470 | 4.5 | 100 |
| $B_{R,L}$ | 20 | 490 | 4.5 | 200 |
| $C_R$ | 20 | 490 | 4.5 | 200 |
| spiral species and handedness | cobalt thickness (nm) | step-out frequency (Hz) | maximum velocity ($\mu m/s$) | |
| $A_R$ | 200 | 11 | 30 | |
| $B_{R,L}$ | 200 | 5 | 75 | |
| $C_R$ | 400 | 10 | 110 | |

Three species of helices were used as the mobile components of the transchiral motors. Their geometry and results are summarized in Table I. The second and third species possessed an additional half turn with a tapering diameter on both ends, to ensure the spiral was contained and would not exit the axial rod. As our micromotors have an overall length of a few millimeters, the weight of the micromotors and the axial rod cannot be overcome by the propulsion of the micromotors. Thus, the micromotors are characterized while near the surface of their environment. The rotation of the spirals induced by the rotating magnetic field also contributes a rolling motion when near the surface.[21] We employ microfluidic channels to constrain the lateral motion of the micromotors, such that the micromotors will only move along their long axis. The micromotor is thus bound by a 12 mm long channel with a square cross-section of 300 x 300 $\mu m^2$.

Before each set of experiments the channels were rinsed with ethanol then sonicated for two minutes, which was repeated with deionized water. The sample was dried then treated with oxygen plasma at 0.2 mBar pressure for two minutes. The channel was placed in a petri dish and filled with a mixture of 1% polysorbate 20 and deionized water. In an experiment, we tested the clockwise and counter-clockwise rotation directions of the magnetic field for ten seconds each. Toward eliminating any possible preferred directionality due to vibration or lithography artifacts, this process was repeated three times. In the first, the entire workspace was rotated 180°, then the transchiral motor was manually rotated 180°, the workspace was then rotated 180° again for the fourth run. For each ten second run, a least squares linear fit was applied to the position of the micromotor, this process yielded eight velocities for a given frequency. Some data points were manually removed if post-processing revealed the micromotor was unable to overcome static friction.

TABLE II. The properties of the transchiral micromotor configurations used in physical experiments.

| transchiral configuration | rod diameter ($\mu m$) | disk diameter ($\mu m$) | spiral combination | expected behavior |
|---|---|---|---|---|
| I | 45 | 155 | $B_L + C_R$ | no motion until 5 Hz |
| II | 45 | 155 | $A_L + B_R$ | reversing direction between 5 and 11 Hz |

FIG. 2. **Quantitative results of individual microspirals in axial rods and transchiral micromotors in a 2 mT rotating magnetic field.** Multiple 10 second runs were made for a given frequency and experimental conditions described in the text. Error bars indicate standard deviation of the velocity of multiple runs. Insets for each plot show scanning electron micrographs of corresponding single constituent spirals and transchiral motors. Species are indicated above the spirals. Red indicates the spirals and axial rod which are coated in magnetic material. Yellow coloring indicates inert resin. Scale bars in A and D are 250 µm. A) Single helix with 100 µm diameter. B) Single helix with 200 µm diameter. C) Single helix with 200 µm diameter and double the magnetic material of the spiral used in B). D) Transchiral motor with two identical spirals with opposing handedness and differing magnetic strengths. E) Transchiral motor with heterogeneous spiral helices.

Transchiral micromotors are fabricated using two-photon lithography in a Nanoscribe Photonic Professional GT using IP-S photoresist (Nanoscribe GmbH). In order to minimize fabrication errors, special care was used to make sure the central rod structure was split at the larger portion (disk), that would anchor the structure to the substrate. To ensure the spirals were not permanently attached to the substrate or the center rod, 2 µm diameter support rods were placed on the bottom of each spiral loop and connected to a base layer that when sliced would be one polymerized layer, approximately 700 nm thick. The support cylinders were broken during the development process, allowing the spirals to rest on either the substrate or rod. Sacrificial structures were printed over the passive rod and disks, preventing the deposition of metal onto these structures. When fabricating the micromotor where each spiral has its own cobalt thickness, the spiral with less material was also given a sacrificial cover that was removed in between cobalt sputtering steps. The micromotors were developed in propylene glycol monomethyl ether acetate (Sigma-Aldrich) then rinsed in isopropyl alcohol. The sample was sputter coated with 30 nm-thick titanium (Leica EM ACE600), 200 or 400 nm-thick cobalt (Kurt Lesker Nano 36), then again with 30 nm-thick titanium. The sample was then coated with a layer of perfluorodecyltrichlorosilane (Sigma Aldrich) by physical vapor deposition. The silane layer was approximated to be one molecular layer, ~ 2 nm. Small deviations in the deposition thicknesses due to equipment or environmental changes can contribute to the variance in the observed speed. The sacrificial structures around the spirals would be removed before use in an experiment. To magnetize the spirals, the micromotor would be placed in an enclosed Gel-Box (Gel-Pak) and magnetized in a 1.8 T magnetic field. Cobalt yielded an intrinsic remanent magnetization of ~ 500 kA/m. Other magnetic materials, such as paramagnetic iron oxide, could be utilized with a stronger magnetic field. This ensured the spirals would always be magnetized perpendicularly to the long axis of the micromotor.

To properly quantify the behavior of individual spiral

FIG. 3. **Transchiral micromotor addressability in a 2 mT rotating magnetic field.** The microchannels are outlined in red and motion is indicated by the white arrow. A) Initial positions with configuration designations. B) After ~ 40 seconds rotating at 4 Hz, configuration I had no net motion and configuration II moved to the center. After an additional ~ 55 seconds, rotating at 7 Hz has reversed the propulsion direction of configuration II and configuration I has translated to the right. (Multimedia view)

species, each was fabricated with the axial, tapered rod and no opposing helix. The quantitative results are given in Fig. 2A-C for species A-C, respectively, and summarized in Table I. The transchiral motor results are summarized in Table II. Figure 2D shows the experimental results of the case illustrated in Fig. 1C. The helices used were species $B_L$ and $C_R$. It was expected that there should be no net motion until species $B_L$ reached its step-out frequency, 5 Hz. However, between 5 and 10 Hz large velocity variance was observed. The peak velocity, 46 µm/s, was observed at the step-out frequency of species $C_R$, 10 Hz. The velocity nonlinearly decayed afterwards. It is important to note this is less than half of the peak velocity of spiral species $C_R$, despite that there should be no translational contribution of spiral species $B_L$. Additional mass of the second spiral and fluid coupling of the spirals could explain the reduction in peak velocity.

The second transchiral motor behavior illustrated in Fig. 1D shows the case of when the spirals do not have equal geometry, including opposing handedness. The transchiral motor was fabricated with helix species $A_L$ and $B_R$, Fig. 2A, B. The results for both rotation directions are shown in Fig. 2E. The helix from species $A_L$ has an opposing handedness that yields propulsion in the negative direction. The peak velocities in the positive and negative directions correspond to the step-out frequency of each respective helix, 5 and 11 Hz.

To provide a demonstration that this method could be used to independently actuate two micromotors in a single workspace, both configurations were placed into two parallel separate microchannels. Their initial positions are shown in Fig. 3A (Multimedia view). The magnetic field was rotated initially at 4 Hz for 40 seconds. Configuration I had negligible net motion while configuration II translated over one body length in the positive x-axis. In the subsequent 54 seconds, configuration I moved in the positive x-axis, and configuration II reversed direction to propel in the negative x direction.

The results from Fig. 2 show high variance in the forward velocity of the micromotors at each input frequency. While multiple runs were completed to ensure that there was no preferred directionality of the micromotor or microchannel, stochastic stiction, vibration, or imperfections in the microstructures could result in a distribution of velocities. As the helices are always rotating, it is possible for them to encounter friction on the tapering disks of the axial rod. If the disk was too small, the spiral would not be contained. A larger disk would yield greater drag and inertia on the mi-cromotor. At a diameter approximately equal to that of the spiral, then the spiral may become hooked on to the circle, and cease rotation. Vibration of the workspace was kept minimal through relatively small magnetic field magnitudes. To minimize stiction, the micromotors were given a monolayer coating of low surface energy fluorosilane, the microchannels were treated with oxygen plasma to ensure a high energy surface that would completely wet, and surfactant was added to the aqueous media.

The proximity to the microchannel walls and axial rod induce additional fluid drag on the micromotors.[37] In general, these wall effects are dependent on the cube of the distance to the wall and become significant within one body length, the diameter of the spiral, of the wall.[38] Within a few microns of contact, the fluid drag can be expected to be approximately 20% greater.[10] The proximity of one helix to another yields the possibility of fluid drag coupling. In this effect, a rotating helix induces a rotational fluid flow. This flow then acts on the second helix, rotating it and yielding a forward propulsion. The separation of micromotors is thus facilitated by the axial rod, as the induced fluid flow decays with the square of the distance from the helix. However, each separation comes with additional structure mass and increases the overall size of the motor and thus the separation distance is an important design parameter of transchiral motors.

The microchannel side walls prevent the rolling of the microswimmer on the channel surface and remove the need of steering the microswimmer during characterization. A single helical spiral is able to steer and reorient by a change in the direction of the rotating magnetic field, which induces a rigid body torque on the helix.[21] Far from a solid boundary, the transchiral motor can be steered similarly since the entire microswimmer has a net magnetization which can be used to reorient the microswimmer.

This methodology theoretically allows an infinite number of spirals in one micromotor, and each could step-out at a different frequency.[36] A special case of the three spiral micromotor is to yield no net motion before the first and after the third step-out frequencies. The spirals need to be designed such that there is no net motion until the first spiral steps out. In addition, the third spiral's behavior after step-out must counter the other two spirals' nonlinear decay. If these conditions are met, there is only net motion in a specific frequency range. A workspace with multiple micromotors with different frequency ranges can then independently drive each motor with no net motion to the other. The size of the micromotor is limited by physical scaling. The magnetic torque scales with the volume of the magnetic material while friction scales with the surface contact area and viscous drag scales with length.[39] Intrinsically stronger magnetic materials and more slippery coatings would allow for further decreasing of the micromotor's size.

In this letter we have presented the fabrication and experimental characterization of transchiral micromotors, swimming microrobots with two magnetic helical structures free to independently rotate. Their translation forces are coupled through an axial tapered rod which restricts their forward motion. We showed two configurations, one which did not have a net propulsion until a critical frequency, and one which had direction reversal at higher frequencies. This work has shown that multiple motion primitives are possible with magnetic micromotors, and that complex, efficient, and submillimeter remote swimming machines are part of the microrobotics paradigm. Future work will focus on the fabrication of micromotors an order of magnitude smaller for use in real-world 3D applications. A constant swimming offset has been shown to compensate for gravity in order to yield no net translation in 3D, though this would require a more complex coupling mechanism, precluding the use of a swimmer that has no net propulsion for a given frequency band.[40]

We would like to thank Frank Thiele and Gunther Richter for assistance and fruitful discussions on cobalt sputtering. J.G., W.W., and M.S. were funded by the Max Planck Society. P.K. was funded by the EPSRC. This project has also received funding from the European Research Council (ERC) under the European Union's Horizon 2020 research and innovation programme (grant agreement 682754 to EL).

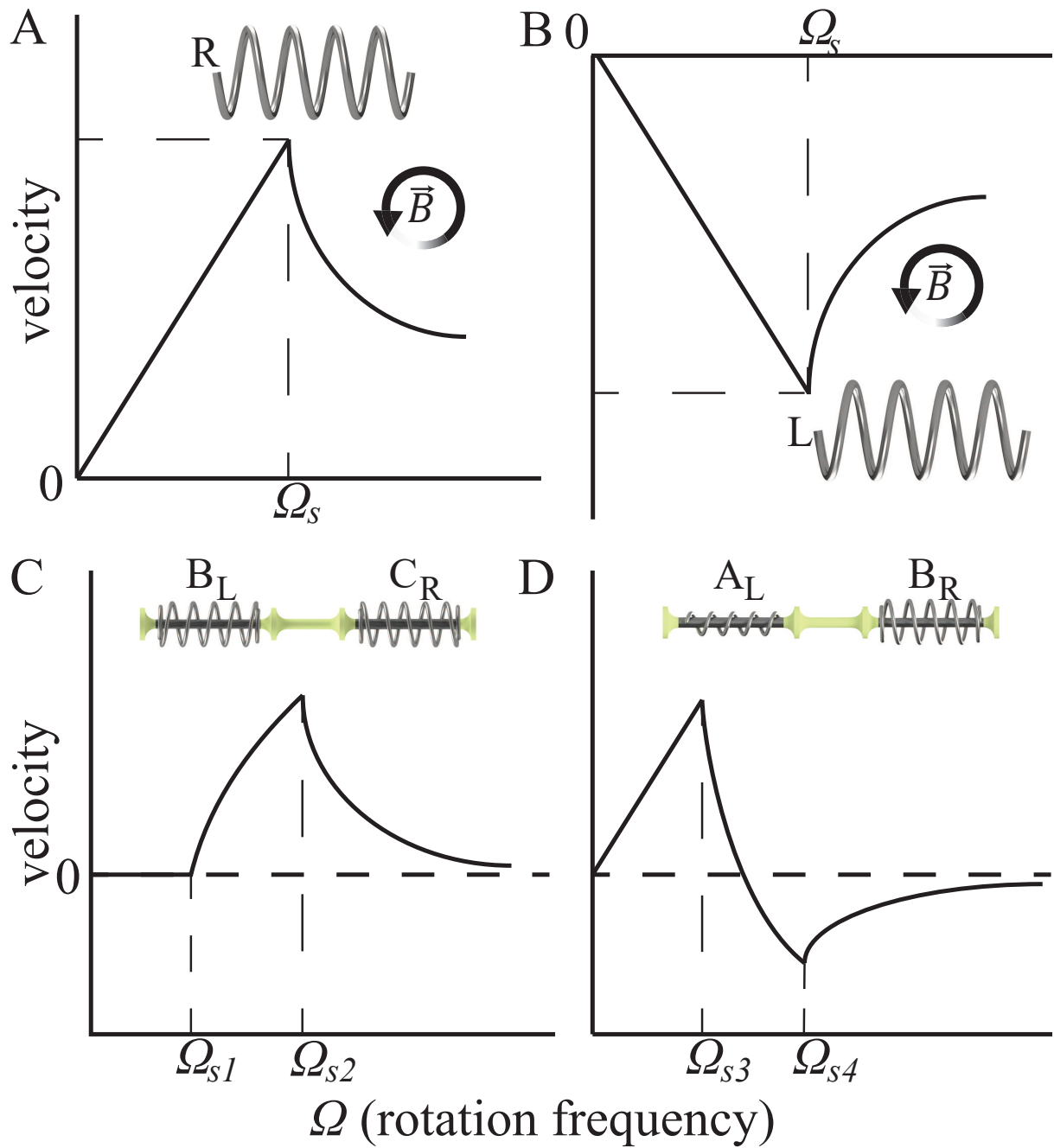

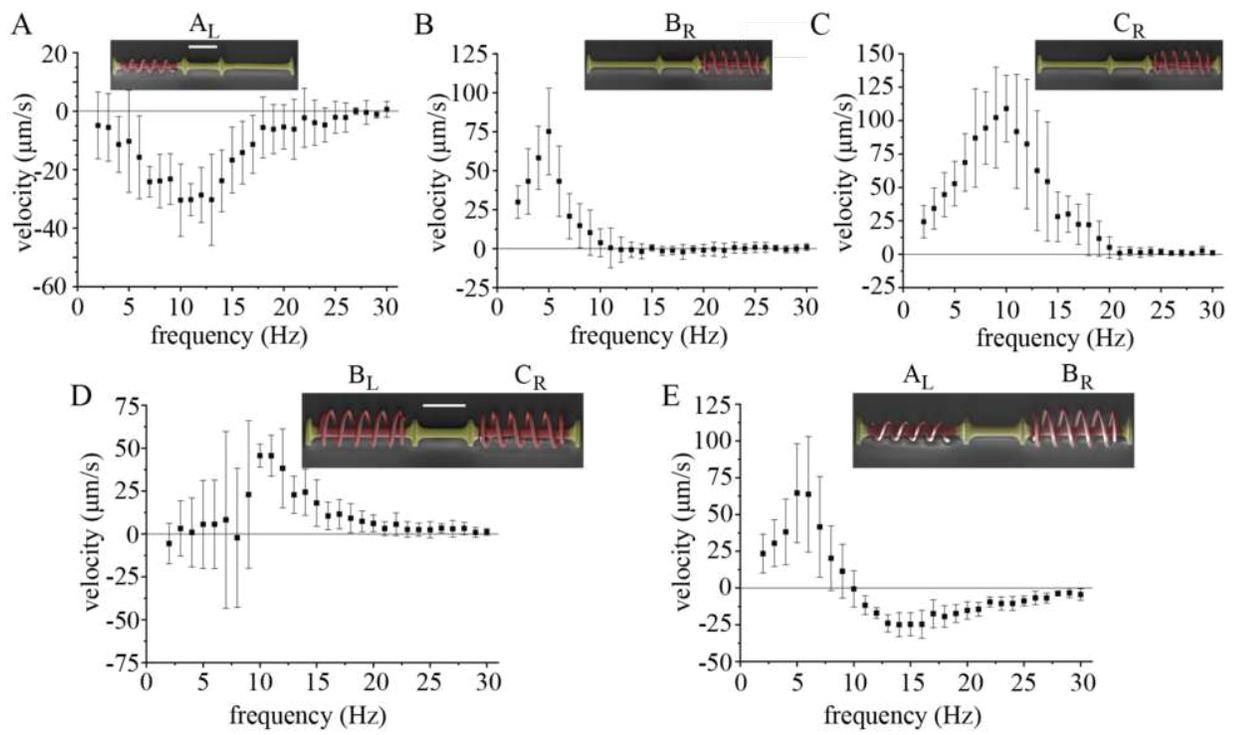

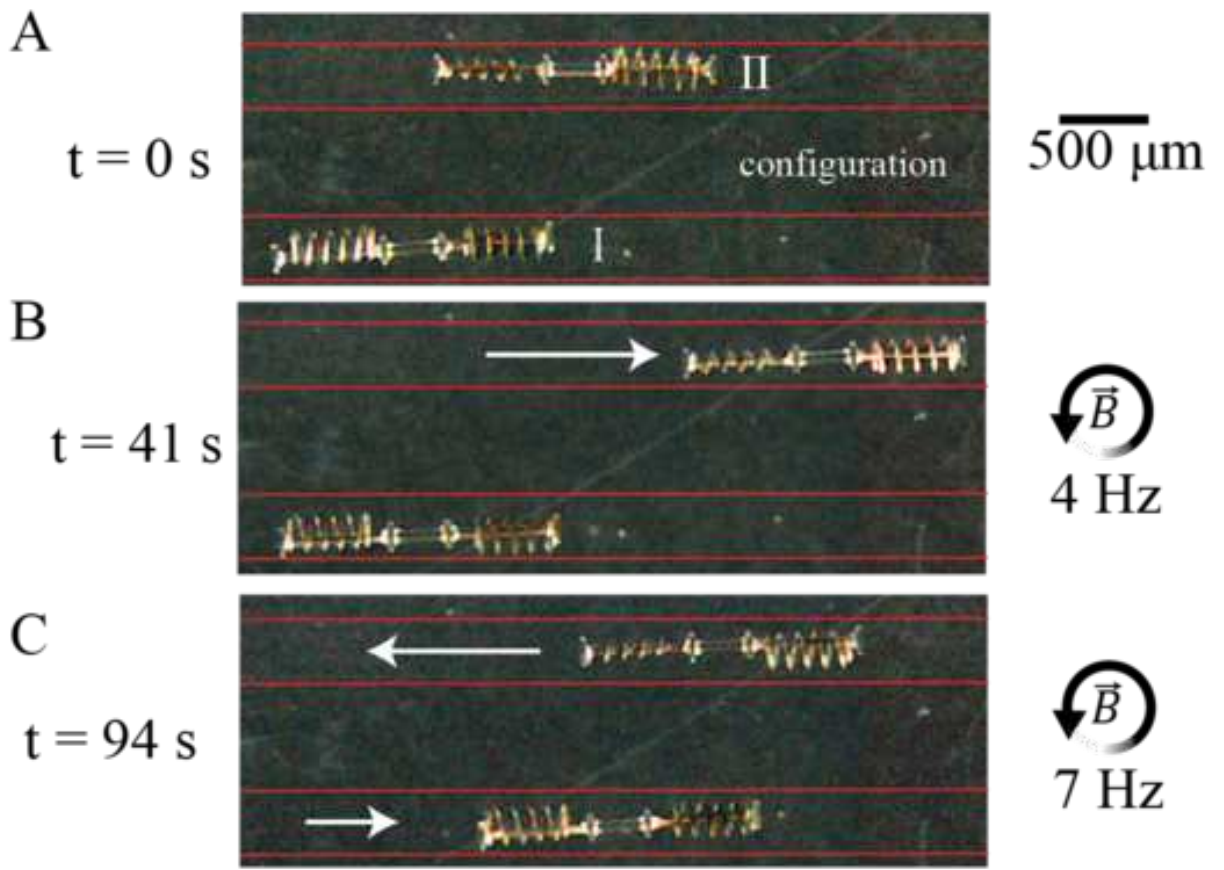